\def\year{2019}\relax
\documentclass[letterpaper]{article} 
\usepackage{aaai19.m} 
\usepackage{times} 
\usepackage{helvet} 
\usepackage{courier} 
\usepackage{url} 
\usepackage{graphicx} 
\usepackage{amsmath,amsthm,amssymb}
\usepackage{mathrsfs,amsmath}
\usepackage[ruled,vlined]{algorithm2e}\usepackage{tikz}
\usetikzlibrary{arrows,%
 petri,%
 topaths,calc,decorations.pathreplacing}
\usepackage{tkz-berge}
\usepackage[position=top]{subfig}
\tikzset{
 font={\fontsize{14pt}{12}\selectfont}}

\renewcommand{\vec}[1]{\boldsymbol{#1}}

\def\calS{\mathcal{S}}
\def\calH{\mathcal{H}}

\DeclareMathOperator*{\argmax}{arg\,max}

\def\newpar{\vspace{-2mm}\paragraph}

\def\ol{\overline}
\newcommand{\coursename}{COST Summer school on Social Choice}

\newcommand{\handout}[5]{
   \renewcommand{\thepage}{#1 \arabic{page}}
   \noindent
   \begin{center}
   \framebox{
      \vbox{
    \hbox to 5.78in { {\bf \coursename}
         \hfill #2 }
       \vspace{4mm}
       \hbox to 5.78in { {\Large \hfill #5  \hfill} }
       \vspace{2mm}
       \hbox to 5.78in { {\it #3 \hfill #4} }
      }
   }
   \end{center}
   \vspace*{4mm}
}

\newenvironment{proof-sketch}{\noindent{\bf Sketch of Proof}\hspace*{1em}}{\qed\bigskip}
\newenvironment{proof-idea}{\noindent{\bf Proof Idea}\hspace*{1em}}{\qed\bigskip}
\newenvironment{proof-of-lemma}[1]{\noindent{\bf Proof of Lemma #1}\hspace*{1em}}{\qed\bigskip}
\newenvironment{proof-attempt}{\noindent{\bf Proof Attempt}\hspace*{1em}}{\qed\bigskip}

\def\ol{\overline}

\def\defeq{\triangleq}

\makeatletter
\def\fnum@figure{{\bf Figure \thefigure}}
\def\fnum@table{{\bf Table \thetable}}
\long\def\@mycaption#1[#2]#3{\addcontentsline{\csname
  ext@#1\endcsname}{#1}{\protect\numberline{\csname
  the#1\endcsname}{\ignorespaces #2}}\par
  \begingroup
    \@parboxrestore
    \small
    \@makecaption{\csname fnum@#1\endcsname}{\ignorespaces #3}\par
  \endgroup}
\def\mycaption{\refstepcounter\@captype \@dblarg{\@mycaption\@captype}}
\makeatother

\newcommand{\mathify}[1]{\ifmmode{#1}\else\mbox{$#1$}\fi}
\newcommand{\bigO}O

\renewcommand{\vec}[1]{{\boldsymbol #1}}

\def\calS{{\cal S}}
\def\calT{{\calT}}

\newcommand{\remove}[1]{{}}

\newtheorem{theorem}{Theorem}
\newtheorem{example}[theorem]{Example}
\newtheorem{lemma}[theorem]{Lemma}
\newtheorem{definition}{Definition}
\newtheorem{observation}[theorem]{Observation}
\newtheorem{proposition}[theorem]{Proposition}
\newtheorem{corollary}[theorem]{Corollary}

\newcount\Comments 
\newcommand{\kibitz}[2]{\ifnum\Comments=1{\color{#1}{#2}}\fi}

\newcommand{\rmr}[1]{\kibitz{red}{[RM: #1]}}

\newcommand{\commentout}[1]{}

\begin{document}
%

\title{Heuristic Voting as Ordinal Dominance Strategies}
\author{Omer Lev$^1$ \and Reshef Meir$^2$ \and Svetlana Obraztsova$^3$ \and Maria Polukarov$^4$\\
$^1$ Ben-Gurion University~~~ \tt{omerlev@bgu.ac.il}\\
$^2$ Technion-Israel Institute of Technology~~~ \tt{reshefm@ie.technion.ac.il}\\
$^3$ Nanyang Technological University~~~ \tt{lana@ntu.edu.sg}\\
$^4$ King's College London~~~ \tt{maria.polukarov@kcl.ac.uk}
}
\maketitle


\begin{abstract}
Decision making under uncertainty is a key component of many AI settings, and in particular of voting scenarios where strategic agents are trying to reach a joint decision. The common approach to handle uncertainty is by maximizing expected utility, which requires a cardinal utility function as well as detailed probabilistic information. However, often such probabilities are not easy to estimate or apply.

To this end, we present a framework that allows ``shades of gray'' of likelihood without probabilities. Specifically, we create a hierarchy of sets of world states based on a prospective poll, with inner sets contain more likely outcomes. This hierarchy of likelihoods allows us to define what we term ordinally-dominated strategies. We use this approach to justify various known voting heuristics as bounded-rational strategies.
\end{abstract}


\section{Introduction}\label{sec:intro}

The question of how agents -- human or artificial -- choose a strategy when facing a choice, has been at the center of attention in artificial intelligence since its inception. Approaches to decision making often rely on two primary components: the \emph{epistemic state} of the agent (her beliefs on how her actions will affect the world), and her \emph{innate preferences} (the utility or cost associated with each outcome). 

In voting scenarios, agents' actions are aggregated to reach a shared result. Voters can make strategic choices once they know what the state of the world is (what other agents are voting), following their own utility function (in most voting settings, an ordinal preference over possible outcomes is assumed). 
This voting decision may either be applied once based on the current beliefs of the voters, or in an iterative fashion so that voters have several opportunities to observe the world and change their action. When the votes of others are unknown, the epistemic state might depend on some prior knowledge and/or signals from the environment. 

The most common way to address this lack of knowledge is to assign probabilities to each state of the world 
and assume that agents each maximize their expected utility over all possible states (see Related Work). 
However, 
%
%
 in many situations human agents may not have the ability to determine precise probabilities of each state of the world, or to act according to them~\cite{TK74,chater2006probabilistic}. There is no reason to believe that in voting scenarios 
people will 
perform differently in this respect. 

\newpar{Voting without Probabilities}
Alternative approaches, focusing on decision making in face of \emph{strict uncertainty} (defined in terms of possible or impossible states) have been formulated and applied in various AI and economic settings~\cite{dow1994nash,boutilier1994toward,Hal97,matt2009dominant}, and more recently, in the context of voting~\cite{CWX11,RE12,MLR14}. The idea at the core of this approach is that, for any given voter, a vote $a'$ \emph{locally dominates} another vote $a$ if $a'$ is at least as good as $a$ for this voter in any voting profile that she considers to be possible. 


In fact, voting behavior can also be defined without any explicit epistemic model. Indeed, several recent papers suggest various heuristics that are specific to a voting rule and/or context. Some of these heuristics have been shown to be empirically consistent with voters' behavior in lab experiments~\cite{laslier2010laboratory}, and others guarantee desirable convergence and/or welfare properties when applied by all voters in a group~\cite{GLRVW13}.

\newpar{Contribution}
We extend the framework suggested in the strict uncertainty papers mentioned above, by \textbf{allowing gradual levels of uncertainty}. 
Specifically, we build on the idea of having a likelihood hierarchy -- a sequence of sets of states of the world, where each next set is a superset of the previous set in the sequence, so that the states in inner sets are considered by the voter to be more likely. 
An \emph{undominated vote} in this setting is one which is not dominated at any level of the hierarchy.

Using this hierarchy of likelihood, we suggest \textbf{an alternative representation for information structures in voting}. 
We show how the relevant information can be boiled down to what we call a \emph{pivot-graph}, which succinctly captures all situations where the voter may be pivotal. We then show that the information structure allows us to justify several existing voting heuristics as rational decisions for an appropriate epistemic model (a specific hierarchy of pivot-graphs). This observation enables us to generalize existing convergence results in the literature on iterative voting, by showing how convergence follows from topological properties of the pivot-graphs.



\newpar{Related Work}\label{relWork}
For an up-to-date coverage of iterative voting, heuristics and uncertainty-based models, see \cite{ME17}. 
In particular, Conitzer et al.~\shortcite{CWX11} consider a voter facing an arbitrary information set, and Reijngoud and Endriss~\shortcite{RE12} study partial information settings where, for example, only the candidates' scores or only the identity of the leader are known. 
Closest to our paper is the \emph{local dominance} model~\cite{MLR14}, in which all voters base their belief on a shared \emph{prospective state}. It's been shown that in an iterative voting setting where voters play possible actions that dominate 
their current action, they are guaranteed to converge to an equilibrium under certain assumptions on the distance metric.

\emph{Voting heuristics} do not explicitly define voters' beliefs; instead, they specify a (typically) simple function that dictates a vote in every given state, aiming to capture realistic voting behaviors~\cite{RE12,GLRVW13}. In particular, some models suggest that a non-pivotal voter either votes truthfully~\cite{DL10} or abstains~\cite{DE10}.


These models stand in contrast with the expected utility models, such as, for example, the \emph{calculus of voting}~\cite{MW93} for a large number of voters, where a voter computes the probability for each action (vote) to be pivotal in every pairwise tie. 
We see our model as a way to capture a similar line of reasoning in identifying the influential ties, albeit without using probabilities. A more fundamental difference with the calculus of voting approach is that the latter assumes a common knowledge of rationality and the preference distribution, from which an equilibrium is derived.

There are also other non-probabilistic models of uncertainty, where two of the most prominent ones are the \emph{possibility theory}~\cite{DFP96} and \emph{Dempster-Shafer theory}~\cite{shafer1976mathematical}. These models attribute a cardinal possibility measure to states and develop calculus rules for belief updates and comparisons. The closest to our work is the \emph{plausibility measure} approach~\cite{Hal97}, that allows for a partial order of plausibility. Our hierarchical ordinal dominance concept is even more strict, and relies on the structure of the problem where uncertainty is essentially about the accuracy of a single point estimate (a poll). 

While we take our distance-based epistemic assumptions from the aforementioned local dominance voting model, an earlier precursor of this idea is the logic for inexact knowledge based on \emph{margin-of-error}~\cite{williamson1992inexact}.

\section{Model}\label{sec:model}

An election is composed of a set $V$ of $n$ voters and a set $C$ of $m$ candidates. 
Each voter~$i\in V$ has a weak preference relation $\succsim^{i}\subseteq C\times C$ over the candidates, that is, for each two candidates $x,y\in C$, $x\succsim_{i}y$ or $y\succsim_{i}x$, and if both are true, they are equivalent. Moreover, the relation is transitive (so, for $x,y,z,\in C$, $x\succsim_{i}y$ and $y\succsim_{i}z$, then $x\succsim_{i}z$). 

The voting rules we shall focus on are the \emph{score-based voting} (SBV) rules. An SBV $(\hat f,A)$ is defined by a set $A\subseteq \mathbb N^m$ of allowed votes, and a function $\hat f:A^{n}\rightarrow C$. For example, the set $A$ under Plurality contains all vectors which have only one non-zero element, which is $1$; Approval allows all binary vectors; Borda allows all permutations of $(0,\ldots,m-1)$; etc. 
We denote by $\vec a = (a_i)_{i\in V}$ the \emph{voting profile}; by $a_i(c)\in \mathbb N$ the absolute score given to $c$ by agent $i$ in vote $a_i\in A$; and by $s_{\vec a}(c)=\sum_{i\in V}a_i(c)$ the total score given to candidate $c$. This creates an aggregated score vector $s$ of size $m$, in which each coordinate corresponds to a different candidate, and its value is $s_{\vec a}(c)$. The winner is $\hat f(s)=\argmax_{c\in C}s(c)$, breaking ties lexicographically.

For each voter $i$, the outcome (and thus, her utility) depends on her own vote, as well as on the state of the world $s=s_{\vec a_{-i}}$, that encompasses the votes of all other participants. We separate these two arguments by writing the outcome function as $f(s,a_i) = \hat f(s+a_i)$. For every state $s$ and any two actions $a_i,b_i\in A$, we write $f(s,a_i) \succ^i f(s,b_i)$ when voter $i$ prefers action $a_i$ over $b_i$ at state $s$.

\begin{example}\label{ex:simple}
There are 100 voters and 5 candidates -- $w,b,c,d,e$ -- using the Plurality voting system. The voters have access to a poll where votes are $\ol s = (29,26,22,17,6)$.
 A voter $i$ currently voting for $b$ sees the state $s=s_{\vec a_{-i}} =(29,\mathbf{25},22,17,6)$. For any action $a'_i$ of $i$, $f(s,a'_i)= w$, which means that voter $i$ is indifferent between her actions. 
\end{example}
 Note that the voters never explicitly reason about the preferences of other individuals -- only about their (aggregated) actions.
 We will return to this example later in the paper. 

\subsection{Information structures}
An \emph{information set} is a set of states $S'\subseteq S$. An \emph{information structure} of agent~$i$ is a collection of information sets $\calS^i=(S^i_{j})_{j=1}^k$, where $S^i_j\subseteq S^i_{j+1}$ for all $j$. That is, each information set contains the sets with a lower index.

An agent does not assign probabilities to states or to information sets, but an intuitive interpretation of the model is that agent~$i$ believes any state in $S^i_j$ to be \emph{substantially more likely} than all states outside $S^i_j$. An information structure can either be shared by all agents, or be agent-specific. 

\begin{example}\label{ex:info}
Consider voter $i$ from Example~\ref{ex:simple} and assume she has an information structure $\calS^{i}=(S^{i}_1,S^{i}_2)$ as in Figure~\ref{fig:information}. In particular, $i$ believes that candidate $c$ may win (as, e.g., $s'=(24,21,26,17,6)\in S^i_2$); however, this is far less likely than a victory of $w$ or $b$, as there is no state $s''\in S^i_1$ where $c$ wins. 
\end{example} 
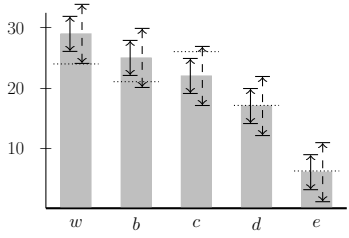
\begin{figure}
\begin{center}
\tikzset{
 font={\fontsize{18pt}{12}\selectfont}}
\begin{tikzpicture}[baseline,scale=0.4,transform shape]
\tikzstyle{sensor}=[draw=black,fill=white,
inner sep=0pt,minimum size=5mm]

\draw[thick] (0,1) -- (10,1);

\node at (1,0.5) {$w$};
\filldraw[black!25!white] (0.5,1.05) rectangle (1.5,6.8);
\draw[,densely dotted] (0.25,5.8) -- (1.75,5.8);

\draw[|<->|] (1-0.2,6.8-0.6) -- (1-0.2,6.8+0.6);
\draw[|<->|,dashed] (1+0.2,6.8-1) -- (1+0.2,6.8+1);

\node at (3,0.5) {$b$};
\filldraw[black!25!white] (2.5,1.05) rectangle (3.5,6);
\draw[,densely dotted] (2.25,5.21) -- (3.75,5.21);

\draw[|<->|] (3-0.2,6-0.6) -- (3-0.2,6+0.6);
\draw[|<->|,dashed] (3+0.2,6-1) -- (3+0.2,6+1);

\node at (5,0.5) {$c$};
\filldraw[black!25!white] (4.5,1.05) rectangle (5.5,5.4);
\draw[,densely dotted] (4.25,6.21) -- (5.75,6.21);

\draw[|<->|] (5-0.2,5.4-0.6) -- (5-0.2,5.4+0.6);
\draw[|<->|,dashed] (5+0.2,5.4-1) -- (5+0.2,5.4+1);

\node at (7,0.5) {$d$};
\filldraw[black!25!white] (6.5,1.05) rectangle (7.5,4.4);
\draw[,densely dotted] (6.25,4.42) -- (7.75,4.42);
\draw[|<->|] (7-0.2,4.4-0.6) -- (7-0.2,4.4+0.6);
\draw[|<->|,dashed] (7+0.2,4.4-1) -- (7+0.2,4.4+1);

\node at (9,0.5) {$e$};
\filldraw[black!25!white] (8.5,1.05) rectangle (9.5,2.2);
\draw[,densely dotted] (8.25,2.24) -- (9.75,2.24);

\draw[|<->|] (9-0.2,2.2-0.6) -- (9-0.2,2.2+0.6);
\draw[|<->|,dashed] (9+0.2,2.2-1) -- (9+0.2,2.2+1);

\draw (0,1) -- (0,7.5);
\node at (0,3) {$-$};
\node at (-1,3) {$10$};
\node at (0,5) {$-$};
\node at (-1,5) {$20$};
\node at (0,7) {$-$};
\node at (-1,7) {$30$};
\end{tikzpicture}
\end{center}
\caption{\label{fig:information} Information structure of voter $i$ at state $s_{\vec a_{-i}} =(29,\mathbf{25},22,17,6)$ (gray pillars). The set $S^{i}_{1}$ contains all states that result from changing the score of any candidate in $s_{\vec a_{-i}}$ by at most 3 votes (solid arrows). The set $S^{i}_{2}$ allows for a variation of 5 votes (dashed arrows). The dotted lines indicate the state $s'$ from Example~\ref{ex:info}.}
\end{figure}

\subsection{Ordinal dominance}
Following \cite{CWX11,RE12,MLR14}, for any information set $S^i_j$ and actions $a,b\in A$, we say that action $a$~~ {\it $S^i_j$-dominates} action $b$ (denoted $a \succ^i_{j} b$) if $f(s,a) \succsim^i f(s,b)$ for \emph{all} $s\in S^i_j$ and $f(s,a) \succ^i f(s,b)$ for \emph{at least one} $s\in S^i_j$. Agent $i$ is \emph{indifferent} between actions $a,b$ at $S^i_j$ (denoted $a \sim^i_{j} b$) if $f(s,a) \sim^i f(s,b)$ for all $s\in S^i_j$. Note that $S^i_j$-dominance is a \emph{partial order} over actions $A$ (transitive, antisymmetric and irreflexive relation). 

\begin{definition}
Action $a$ \emph{ordinally dominates} action $b$ (in structure $\calS^i =(S^i_j)_{j\in [k]}$) if
there is some $j \in [k]$ such that action $a$ $S^i_j$-dominates action $b$.
\end{definition}

The next lemma guarantees that it is not possible that $a \succ_j b$ and $b \succ_{j'} a$ for some $j'\neq j$.
\begin{lemma}Ordinal dominance is a partial order.
\end{lemma}
\begin{proof}
\underline{Transitivity}: Suppose action $a$ ordinally dominates action $b$ and $b$ ordinally dominates action $c$, due to $S^i_j$ and $S^i_{j'}$, respectively. W.l.o.g. $j'\leq j$, then $S^i_{j'} \subseteq S^i_{j}$. There is a state $s'\in S^i_{j'}$ where $f(s',b) \succ^i f(s',c)$, and since $s'\in S^i_{j'}\subseteq S^i_{j}$, we also have $f(s',a) \succeq^i f(s',b)$, and so $f(s',a)\succ^i f(s',c)$. Similarly, for any $s\in S^i_{j'}$, $f(s,a) \succeq^i f(s,b) \succeq^i f(s,c)$. Thus $a \succ^i_{j'} c$ which means that $a$ ordinally dominates $c$. 

\underline{Antisymmetry}: Suppose action $a$ ordinally dominates action $b$ due to $S^i_j$. For every $j'\leq j$, there cannot be a state $s\in S^i_{j'}\subseteq S^i_j$ where $f(s,b) \succ^i f(s,a)$. Similarly, for any $j'>j$, there is a state $s'\in S^i_j \subseteq S^i_{j'}$ where $f(s',b) \prec^i f(s',a)$. Thus, $b$ does not $S^i_{j'}$-dominates $a$. Since this is true for any $j'$, $b$ does not ordinally dominate $a$. 
\end{proof}

\def\celcius{^\circ}
\subsection{Distance-based uncertainty}
Following Meir et al.~\shortcite{MLR14}, we consider the following way to derive information sets and information structures. 
 Given a metric $d:S\times S\rightarrow \mathbb [0,1]$ and a parameter $r\in [0,1]$, every state $s\in S$ explicitly defines an information set $S_{d,r}(s)=\{s' : d(s,s')\leq r\}$. In general, the metric $d$ can be completely arbitrary and the induced set is meaningless.\footnote{In fact, \emph{any} set $S'\subseteq S$ can be derived from $s$ for some carefully designed metric $d$.} However, in the context of voting there are several natural metrics: For example, $d(s,s')$ may reflect what fraction of votes has changed between $s$ and $s'$. 
 In Meir et al.~\shortcite{MLR14} and Meir~\shortcite{Mei15}, the distances between candidate score vectors were defined by different $\ell$-norms and the Earth Mover distance (EMD), 
which is essentially the $\ell_1$ norm with the additional constraint that the total number of votes remains the same. Thus, $S_{d,r}(s)$ may reflect a range of possible candidates' scores given a poll or a current state $s$. 


\begin{figure}
%
%
\centering
\tikzset{
  font={\fontsize{24pt}{12}\selectfont}}
%
%
%
%
%
%
%
%
%
%
%
%
\tikzset{
  font={\fontsize{18pt}{12}\selectfont}}
\begin{tikzpicture}[baseline,scale=0.45,transform shape]
\filldraw[black!5!white] (0,2) rectangle (6,10);
\filldraw[black!15!white] (0,4) rectangle (5.8,10);
\filldraw[black!28!white] (0,6) rectangle (5.6,10);
\filldraw[black!40!white] (0,8) rectangle (5.4,10);
\draw[thin] (0,2) rectangle (6,10);
\draw[thin] (0,4) rectangle (5.8,10);
\draw[thin] (0,6) rectangle (5.6,10);
\draw[thin] (0,8) rectangle (5.4,10);

\node at (3,9.6) {$(30,26,22,17,5)$};
\node at (2.7,8.8) {$\ol s=(29,26,22,17,6)$};
\node at (3,8.2) {$\ldots$};
\node at (3,7.6) {$(26,26,24,16,8)$};
\node at (3,6.8) {$(26,27,23,18,6)$};
\node at (3,6.2) {$\ldots$};
\node at (3,5.6) {$(30,30,22,12,6)$};
\node at (3,4.8) {$(21,26,30,17,6)$};
\node at (3,4.2) {$\ldots$};
\node at (3,3.6) {$(29,26,34,5,6)$};
\node at (2.7,2.8) {$s'=(18,27,22,27,6)$};
\node at (3,2.2) {$\ldots$};
\node at (3,1.6) {$(0,0,0,100,0)$};
\node at (3,0.8) {$(20,20,15,15,30)$};
\node at (3,0.2) {$\ldots$};

\draw [decorate,decoration={brace,amplitude=4pt},xshift=0pt,yshift=0pt] (6.1,10) -- (6.1,4) node [black,midway,xshift=1cm] { $S_3$};
\end{tikzpicture}~~
\begin{tikzpicture}[baseline,scale=0.45,transform shape]

\tikzstyle{EdgeStyle}=[thin]
\node at (11,9) {$H_1$};
\Vertex[x=7,y=9]{d}
\Vertex[x=8,y=9.6]{b}
\Vertex[x=8,y=8.4]{c}
\Vertex[x=9,y=9]{w}
\Vertex[x=10,y=9]{e}

\node at (9,7) {$H_2$};
\Vertex[x=10,y=7]{d}
\Vertex[x=11,y=7.6]{b}
\Vertex[x=11,y=6.4]{c}
\Vertex[x=12,y=7]{w}
\Vertex[x=13,y=7]{e}
\Edge(w)(b)

\node at (11,5) {$H_3$};
\Vertex[x=7,y=5]{d}
\Vertex[x=8,y=5.6]{b}
\Vertex[x=8,y=4.4]{c}
\Vertex[x=9,y=5]{w}
\Vertex[x=10,y=5]{e}
\Edge(w)(b)
\Edge(w)(c)
\Edge(b)(c)

\node at (9,3) {$H_4$};
\Vertex[x=10,y=3]{d}
\Vertex[x=11,y=3.6]{b}
\Vertex[x=11,y=2.4]{c}
\Vertex[x=12,y=3]{w}
\Vertex[x=13,y=3]{e}
\Edge(w)(d)
\Edge(w)(b)
\Edge(w)(c)
\Edge(b)(c)
\Edge(b)(d)
\Edge(c)(d)
\Edge(w)(e)
\tikzstyle{EdgeStyle}=[thin,bend left]
\Edge(b)(e)
\tikzstyle{EdgeStyle}=[thin,bend right]
\Edge(c)(e)
\end{tikzpicture}

%
\caption{\label{fig:voting}A schematic example of the information structure $\calS_{EMD,\vec r}(\ol s)$, and the induced pivot graph structure $\calH_{EMD,\vec r}(\ol s)$. E.g., graph $H_4$ contains the edge $(b,d)$ due to state $s'\in S_4$. States below $S_4$ are considered impossible. All the pivot graphs are upward closed w.r.t. the lexicographic order on $C$, but they are not always a clique ($(d,e)\notin H_{4}$).
}
\end{figure}
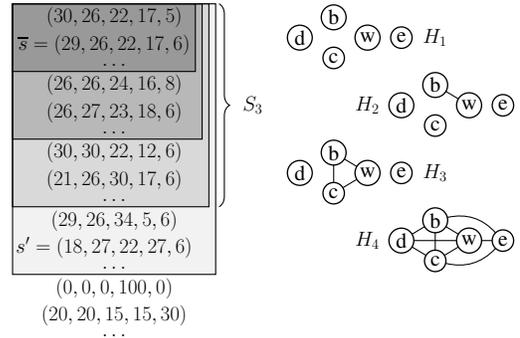

\section{Pivot graphs}
A pair of actions $(a',a'')$ is \emph{pivotal} for a pair of candidates $c',c''\in C$ in state $s\in S$, if $f(s,a')=c'$ and $f(s,a'')=c''$. 
An agent $i$ is pivotal for the pair of candidates $c',c''\in C$ in information set $S^i_j$, if there are $s\in S^i_j$ and actions $a'_i,a''_i\in A$ that are pivotal for $c',c''$ in $s$. 

Information set $S^{i}_j$ then induces a \emph{pivot-graph} $H^i_j=(C,E)$, which contains a vertex for every candidate, and an edge $(c',c'')$ if agent $i$ is pivotal for the pair $c',c''$ in $S^i_j$. 

Every information structure $\mathcal{S}^i$ induces a \emph{pivot graph structure} $\mathcal{H}^i=(H^i_{j})_{j=1}^k$, where each $H^i_j$ is a subgraph of $H^i_{j+1}$ (since adding more states can only add edges to the graph). The set $\mathscr{H}(C)$ contains all pivot graph structures. 


%

\subsection{Epistemic models and OD Equilibrium}
 An \emph{epistemic model} of agent~$i$ maps any state $s$ to an information structure $\calS^i(s)=(S^i_1(s),S^i_2(s),\ldots,S^i_k(s))$, and thus also to a pivot graph structure $\calH^i(s)$. 

 
We define the set $OD_{\succsim^i}(\calS,a)$ that contains all actions that ordinally dominate $a$ in $\calS$ according to preferences $\succsim^i$, and a set $UOD(\calS,a)$ containing all actions $a'$ that ordinally dominate $a$ but are not ordinally dominated themselves. Naturally, this leads to a definition of an OD-equilibrium -- when for every agent $OD_{\succsim^i}(\calS^i(s_{\vec a_{-i}}),a_i)$ is empty (and hence, $UOD_{\succsim^i}(\calS^i(s_{\vec a_{-i}}),a_i)=\emptyset$ too).

\begin{observation}
For a ``full information'' epistemic model where $\calS^i(s)=\left(\{s\}\right)$, the set $OD(\calS^i(s),a_i)$ coincides with the set of better-responses to $(s,a_i)$; the set $UOD(\calS^i(s),a_i)$ coincides with best-responses; and OD equilibrium coincides with a pure strategy Nash equilibrium.
\end{observation}

An epistemic model is \emph{cliqued} if its mapping to a pivot graph structure $\calH(s)$ at every state $s\in S$, $H_j(s)$, is a clique. The epistemic model is \emph{upward closed} if the pivot graph structure $\calH(s)$ at every state $s$ has an order $L$ over candidates such that if $(c,c')\in H_j(s)$ and $c''>_L c'$ then $(c,c'')\in H_j(s)$. Note that any cliqued epistemic model is upward closed (where $L$ may be an arbitrary order where all candidates in $H_j(s)$ precede all others).	More generally, $L$ can be roughly thought of as an order of likelihood of states. For simplicity of notation, we denote both the pivot graph structure and the epistemic model (which is a function from states to structures) by $\calH$.


Distance metrics provide us with a simple way to define an information structure: given a metric $d$ and an increasing sequence of distances $\vec r = (r_1,r_2,\ldots,r_k)$, we get an epistemic model $\calH_{d,\vec r}(s) = (H_{d,r_1}(s),H_{d,r_2}(s),\ldots,H_{d,r_k}(s))$ that is induced by $\calS_{d,\vec r}(s)$. 
We later show how the topological properties of $\calH_{d,\vec r}(s)$ are related to the properties of the metric $d$.

\begin{example}\label{ex:voting} We expand Example~\ref{ex:simple} where candidates' scores (in $\%$ of total) are $(29,26,22,17,6)$. We do not specify the number of voters in the poll. We consider a voter with a concentric information structure, based on the radii $\vec r= \{1\%,3\%,7\%,17\%\}$ and the EMD metric. These information sets induce pivot-graphs as illustrated in Figure~\ref{fig:voting}. 
\end{example}


In Example~\ref{ex:voting}, consider a Plurality voter whose preferences are $e\succ d\succ c\succ b\succ w$. Then the action ``$c$'' (a shorthand for $(0,0,1,0,0)$) ordinally dominates action ``$e$'' (due to $H_3$) and ``$b$'' ordinally dominates everything else due to $H_2$. 

\subsection{Sharp Pivot Property in large populations}
Note that structures $\calS^i$ and $\calH^i$ are two different ways to represent the information of an agent. In general, $\calH^i$ contains less information than $\calS^i$, since there may be states when some action pairs are pivotal, whereas others are not. 

Yet, it seems plausible that in a large population, a voter is unlikely to make such fine distinctions about the possible outcomes: voters do not know the exact score of each candidate, but only have a rough idea of what it is (each candidate's share of the votes). As a result, it is reasonable to assume that if a voter considers herself pivotal in some possible tie, she will consider \emph{any} change in her vote as possibly pivotal. We capture this property in the following formal definition.


\begin{definition}[Sharp Pivot Property (SPP)]\label{def:pivot}
An information structure $\calS^i$ satisfies the \emph{Sharp Pivot Property} if:
for all $j $ and all $c',c''\in C$, an edge $(c',c'')\in H^i_j$ entails that any pair of actions $(a'_i,a''_i)$ such that $a'_i(c')-a'_i(c'') > a''_i(c')-a''_i(c'')$ is pivotal for $(c',c'')$ in $S^i_j$. 
%
%
\end{definition}
That is, if there is some action pair that makes $c'$ the winner instead of $c''$, then \emph{any move} that increases the gap in favor of $c'$ might make $c'$ beat $c''$ and become the winner (in some $s\in S^i_j$). 

 Working with pivot-graph structures is much more convenient than working with arbitrary information sets, and their meaning in the context of voting is clear. We will assume throughout the paper that all information structures have SPP, 
which means that $\calH^i$ contains all the relevant information in $\calS^i$.

\newpar{Justifying SPP}
While SPP is plausible, we would like to show that at least in some cases it provably holds. The argument is a rather fine one, that relies on viewing the (finite) poll as approximating some underlying real-valued distribution $p\in \Delta(C)$. Distribution $p$ defines a unique score vector $s_p^n$ for every population size $n$ (using some fixed rounding of $p\cdot n$), and we argue that for a sufficiently large population $n$, the information structure $\calS_{d,\vec r}(s_p^n)$ satisfies SPP. 

Intuitively, consider the leader $w$ in Example~\ref{ex:voting} with the Plurality rule, some other candidate (say $b$), and some information set $S_{d, r_j}(s)$. If $s(w)-s(b)$ is much lower than $r_j\cdot n$, 
then any move where the voter deserts $w$ and/or joins $b$ might be pivotal for $(b,w)$. In contrast, if $s(w)-s(b)$ is much higher than $r_j \cdot n$, then $b$ can never win or even tie with $w$. 
Only if $s(w)-s(b) \cong r_j \cdot n$, then moving from $w$ to $b$ makes $b$ the winner, but (say) moving from $d$ to $b$ does not. This extreme case becomes unlikely as the population gets larger. 

\begin{theorem}
For the EMD metric $d$, any SBV, and any radii vector $\vec r$, the following holds for almost all\footnote{Except a 0-measure set of distributions.} distributions $p\in \Delta(C)$: There is $n_0$ such that for all $n>n_0$, the information structure $\calS_{d,\vec r}(s^n_p)$ satisfies SPP.
\end{theorem}

\subsection{Computing dominance relations}
We show that strategies can be efficiently compared according to ordinal dominance.
\begin{proposition}
Given a pivot graph structure $\calH^i=(H^i_1,\ldots,H^i_k)$ and any SBV $(\hat f,A)$, voter $i$ can check in time $O(m^2k)$ if vote $a'_i\in A$ ordinally dominates vote $a_i\in A$.
\end{proposition} 

\def\defeq{\leftarrow} 
\def\diff{\text{diff}}
\def\sign{\text{sign}}
\def\pivot{\text{pivot}}
\def\dom{\text{dom}}
\def\effect{\text{effect}}
\def\safe{\text{safe}}
Intuitively, Algorithm~\ref{alg:OD} checks (for each uncertainty level $j$), whether the new vote $a'_i$ is ``safe'' (not worse than $a_i$ in any possible tie), and whether it is ``pivotal'' (better than $a_i$ in at least one tie). 

$I[X]\in \{-1,1\}$ is an indicator variable for statement $X$, and we use $a_{i}(c)$ to indicate candidate $c$'s score when voter $i$ vote is $a_{i}$. \rmr{I don't understand this notation: (i.e., $(\vec a_{-i},a_{i})_{c}$).}
The complexity of checking whether a given vote $a_i$ is \emph{undominated} is left as an open question.

\begin{algorithm}[h]
\begin{small}
\For{$c,c'\in C$} {
$\diff(c,c') \defeq a'_i(c)+a_i(c')-a_i(c)-a'_i(c')$\;
$\effect(c,c') \defeq \sign(\diff(c,c')\cdot I[c\succ^i c'])$\; 
}
\For{$j\leq k$} {
 $\safe(j) \defeq \min_{(c,c')\in H_j^i}\effect(c,c')$\;
	 $\pivot(j) \defeq \max_{(c,c')\in H_j^i}\effect(c,c')$\;
	 $\dom(j) \defeq I[\pivot(j)+\safe(j)\geq 1]$\;
	}
\If{$\exists j\leq k$ s.t. $\dom(j)=1$}{
\KwRet TRUE}\Else{
\KwRet FALSE}
\caption{\small{\textsc{OD}($a'_i,a_i\in A,\succsim^{i},\calH^i\in \mathscr{H}$)}}
\label{alg:OD}
\end{small}
\end{algorithm}

 \begin{proof}
Suppose $a'$ ordinally dominates $a$. Then there is some level $j\leq k$ such that $a' \succ^i_j a$. 
This means that for any pair of candidates $(c,c')$ that can be tied in $H^i_j$, either $c$ is preferred to $c'$ and $a'$ weakly reduced $c'$'s score, or $c'$ is preferred to $c$ and $a'$ weakly adds to $c'$'s score (thus $\effect(c,c')\geq 0$). Hence, in particular $\safe(j)\geq 0$. In addition, there must be a pair for which the gain is strict, and $\effect(c,c')=1$, which means $\pivot(j)=1$. In total, $\dom(j)\geq 1+0 =1$ so the algorithm returns TRUE.

Otherwise, in every level $j$, either $a'_i,a_i$ have the same outcome in all states, or there is a pair $(c,c')\in H^i_j$ such that $f(s,a_i)=c,f(s,a'_i)=c'$, and $c\succ^i c'$.

In the latter case, since $f$ is a scoring rule this means that $a'_i(c)-a_i(c)< a'_i(c')-a_i(c')$, i.e. that $c'$ gained strictly more score than $c$ when changing from $a_i$ to $a'_i$. Thus $\diff(c,c') = a'_i(c)+a_i(c')-a_i(c)-a'_i(c')<0$, and $\effect(c,c')= -1$. The algorithm then computes $\safe(j) = -1$. Therefore $\dom(j)\leq 1-1 = 0$.

In the first case, $\effect(c,c')=0$ for all pairs, and thus $\safe(j)=\pivot(j)=0$, and $\dom(j)=0$. 
\end{proof}

\rmr{omitted: To recap, our model differs from standard (strict uncertainty) epistemic game theory models in two important ways: it does not (necessarily) assume that beliefs form a partition of the state space; and agents do not attribute explicit beliefs, preferences, or rationality to other agents. Rather, our agents each act as bounded rational individual decision makers, reacting to an uncertain but static world. }

\section{Justifying Voting Heuristics with OD}\label{sec:heuristics}

Many heuristics have been suggested to analyze how voters behave and change their vote. 
Most heuristics are derived from a single ``prospective state'' $s$, which is assumed to be the current voting profile or poll. Formally, a \emph{set heuristics} is a function $h:S\times A \rightarrow 2^A$ that maps the prospective state and the current action to a set of new possible actions. We say that $h$ is a \emph{point heuristics} if $|h(s,a)|\leq 1$ for every $s,a$. To be consistent with previous definitions, we always omit $a$ from the set $h(s,a)$, and assume that when $h(s,a)=\emptyset$ the voter simply keeps her current vote. 


\begin{definition} We say that an epistemic model $\calH$ \emph{justifies} heuristic $h$, 
if for any state $s\in S$ and current action $a\in A$: 
 (I) $h(s,a)=\emptyset$ if and only if $UOD(\calS(s),a)=\emptyset$; and (II) $h(s,a)\subseteq UOD(\calS(s),a)$.
$\calH$ \emph{strongly justifies} $h$ if (II) holds with equality.
\end{definition}
This means that the heuristic $h$ only recommends undominated ordinal-dominance moves under the epistemic model $\calH$, and only keeps the current action if no such move exists. 


As a simple example, consider the Plurality rule and the heuristic $h^{not-last}(s,a)$ that is empty except when action $a$ is the least preferred candidate $\hat a_i$, and then it moves to an arbitrary other candidate. 
Consider the epistemic model $\calH^{all}(s) = (H^{all}_1)$ where $H^{all}_1$ is the complete graph. 

\begin{observation} $\calH^{all}$ strongly justifies $h^{not-last}$.
\end{observation}
This is since (I) suppose that $a\neq \hat a_i$. Then no candidate ordinally dominates $a$ and thus $UOD(\calH^{all}(s),a)=\emptyset=h^{not-last}(s,a)$; (II) when $a=\hat a_i$, any other candidate $c$ is undominated but globally dominates $a$ (since there is a possible state where $i$ is pivotal for $c$ against $a$), in which case $UOD(\calH^{all}(s),a)=C\setminus\{\hat a_i\}=h^{not-last}(s,a)$.


\subsection{Local dominance}\label{sec:h_LD}
Local dominance~\cite{MLR14} heuristic with metric $d$ and parameter $r$ explicitly define a set $S_{d,r}(s)=\{s' : d(s,s')\leq r\}$. The heuristic action $h^{LD}_{d,r}(s,a_i)$ is defined for the Plurality rule as follows: Let $D\subseteq C$ be the set of candidates that $S_{d,r}(s)$-dominate $a_i$; If $D$ is non-empty, then vote for the most preferred candidate in $D$.


We define an epistemic model $\calH^{LD}_{d,r}$ where $\calH^{LD}(s)_{d,r}$ contains a single pivot graph $H_1$ which is the pivot graph induced by $S_{d,r}(s)$. Note that our definition applies for \emph{any voting rule}, unlike the one in Meir et al.~\shortcite{MLR14}. In Plurality, $\calH^{LD}_{d,r}$ justifies $h^{LD}_{d,r}$ (straightforward proof omitted due to space constraints).



\paragraph{Truth/lazy-bias} 
Denote the top candidate of $i$ by $q_i\in C$, and denote by $\bot$ an ``abstain'' action that adds no score to candidates.
We adopt the suggested variations in Dutta and Laslier~\shortcite{DL10} and Desmedt and Elkind~\shortcite{DE10}, where the voter prefers the truthful/abstain action if this does no affect the outcome. However, this na\"ive modification alone may lead to unreasonable behaviors, e.g., where no-one votes~\cite{elkind2015equilibria}, even under full information.
\footnote{To be completely formal, the preference relation $\succsim^i$ has to be extended to preferences over pairs (winner,action). See supplementary material for details.} 

For $r_2>r_1$, the ``truth bias'' heuristics $h^{LD+TB}_{d,r_1,r_2}(s,a_i)$ is as follows~\cite{MLR14}: (1) perform a local-dominance move at radius $r_{1}$, if exists. If such move does not exist, $i$ examines if $f(s',a_i) \succ^i f(s',q_i)$ for some $s'\in S_{d,r_2}(s)$. (2a) If so, agent~$i$ keeps the current vote $a_i$, (2b) otherwise, $i$ moves to $q_i$.

While the behavior seems to maintain the reason behind truth bias, the definition of $h$ is cumbersome. Instead, we can use $r_1,r_2$ to define an epistemic model $\calH^{LD+TB}_{d,r_1,r_2}$ as follows. We let $H_1(s)$ be as in $\calH^{LD}_{d,r_1}$ above. 
We similarly compute $H_{2}(s)$ from $S_{d,r_2}(s)$, but taking only edges between the current vote $a_i$ and candidates less preferred than $a_i$. Let $\calH = \calH^{LD+TB}_{d,r_1,r_2}(s) = (H_1(s),H_1(s) \cup H_2(s))$. 
\begin{proposition} $\calH^{LD+TB}_{d,r_1,r_2}$ justifies $h^{LD+TB}_{d,r_1,r_2}$ in Plurality.
\end{proposition}
\begin{proof}
First, if $H_1(s)$ is nonempty (at least one tie) then $h^{LD+TB}_{d,r_1,r_2}(s,a_i)= a^*_i \in UOD(\calH(s),a_i)$ as in a standard LD move. Otherwise, there are two cases. 

If $H_2(s)$ contains some edge $(a_i,b)$, then by SPP for any $a'\neq a_i$ there is a state $s'\in S_{d,r_2}(s)$ where $f(s',a_i)=a_i$ and $f(s',a')=b$ (think of $s'$ as state where a single additional vote for $a_i$ is critical). Since $a_{i}$ is preferred to $b$ by the definition of $H_{2}(s)$, we conclude that no candidate $a'$ dominates $a_i$ in $H_2(s)$ (thus $UOD(\calH(s),a_i)=\emptyset$); and that $f(s',a_i) =a_i \succ^i b= f(s',q_i)$ (thus $h^{LD+TB}_{d,r_1,r_2}(s,a_i)=\emptyset$). 

In the second case, there is no such edge, then $H_2(s)$ is empty. This means that no action of $i$ can change the outcome whatsoever, and thus by the slight truth-bias $q_i$ is strictly preferred to any other action. In particular, it ordinally dominates $a_i$ and is undominated so $UOD(\calH(s),a_i) = \{q_i\}$.
Finally, since $i$ is non-pivotal then in particular there is no state $s'\in S_{d,r_2}(s)$ such that $f(s',a_i) \succ^i f(s',q_i)$. Thus $h^{LD+TB}_{d,r_1,r_2}(s,a_i)=q_i \in UOD(\calH(s),a_i)$, as required. 
\end{proof}

The statement for lazy-bias is similar, and uses the same information structure but with a slight preference to abstain instead of voting truthfully. 

\subsection{$T$-pragmatist} The $T$-pragmatist (point) heuristic \cite{brams1978approval,RE12} considers the leading $T$ candidates in $s$ (denoted $\mathbf T$), and sets a new action $a'_i=h^{T-prag}(s,a_i)$ where $a'_i$ is identical to $a_i$ except the favorite candidate in $\mathbf T$ is given maximal score (in any SBV). E.g., in Example~\ref{ex:simple}, if $e\succ^i c\succ^i b \succ^i w \succ^i d$, then $h^{2\!-\!prag}(s,a_i)\!=\!b$ and $h^{3\!-\!prag}(s,a_i)\!=\!h^{4\!-\!prag}(s,a_i)\!=\!c$.


\rmr{comment: it is less desirable that the beliefs depend on the preferences. If we do not like this, we need to consider changing the heuristics. can we define a different graph structure that is is independent of preferences and induces an alternative TB heuristics (and also k-pragmatist)?}

Consider a single-level epistemic model $i$ $\calH^{T,i-star}(s)$, which contains a star graph, in which the center node is the most preferred candidate by voter $i$ in the top $T$ candidates, and it is tied with all other $T-1$ candidates in the top $T$.

\rmr{add non-myopic voting?}

It is possible to show that (I) $h^{T-prag}(s,a_i)=\emptyset \iff OD(\calH^{T,i-star}(s),a_i)=\emptyset$; and (II) $h^{T-prag}(s,a_i) \subseteq OD(\calH^{T,i-star}(s),a_i)$.
This shows a connection between the heuristic and the epistemic model, but it is not a sufficient justification since $h^{T-prag}(s,a_i)$ may be dominated. A closer look reveals that the actions dominating it are quite plausible: ranking the other candidates in $\vec T$ at the bottom can only benefit the voter! We conclude that the T-pragmatist heuristic could be improved. 

We define the $h^{T*}$ heuristic similarly to $h^{T-prag}$, with the following difference: all other candidates in $\vec T$ \emph{get minimal score} (i.e., ranked at the bottom of $a'_i$) while maintaining the same order among themselves as in $a_i$.
\begin{proposition}
$\calH^{T,i-star}$ strongly justifies $h^{T*}$ in any SBV. 
\end{proposition}
The proof is given in the supplementary material.

\subsection{Leader Rule (Approval voting)} Assume candidates $c_1,\ldots,c_m$ are sorted in decreasing score order in a state $s$. In Approval voting the allowed actions are $A=2^C$. The \emph{Leader rule}~\cite{Las09} $a'=h^{LR}(s,a_i)$ is a strategy approving all candidates strictly preferred to the leader of $s$, and approves the leader of $s$ (candidate $c_1$) if and only if it is preferred to the runner-up $c_2$ (i.e., exactly one of $c_1,c_2$ is being approved in $a'$). 

We consider the epistemic model where $\calH^{LR}(s)$ consists of two nested pivot graphs. The inner graph $H_1$ contains a single edge between $c_1$ and $c_2$. The outer graph $H_2$ is a star connecting $c_1$ to all candidates. 

\begin{proposition} $a'=h^{LR}(s,a_i)$ ordinally dominates all other actions according to $\calH^{LR}$. In particular, $\calH^{LR}$ strongly justifies $h^{LR}$.
\end{proposition}
\begin{proof}
Let $a''$ be any alternative vote to $a'$. We will show that $a'$ dominates $a''$ in at least one of the tie graphs $H_1$ or $H_2$. 

Consider $a''$ that differs from $a'$ on (at least) $c_1$ or $c_2$ or both. On the graph $H_1$, the voter is pivotal for $c_1,c_2$ and thus there is a state $s$ where $f(s,a'')=c_2 \prec^{i} c_1 = f(s,a')$, or $f(s,a'')=c_1 \prec^{i} c_2 = f(s,a')$. Thus $a'$ dominates $a''$ on $H_1$. 

Next, consider $a''$ that approves $c_1,c_2$ iff $a'$ approves them, but differs in (at least) some other candidate $c'$. If $c_1\succ c'$, $c'$ is not approved in $a'$ and thus approved in $a''$ (this is regardless of whether $c_1$ is approved). Since there is a state $s$ in $H_2$ where $c_{1}$ and $c'$ are tied, $f(s,a'')=c' \prec c_1 = f(s,a')$. If $c_1\prec c'$, $c'$ is approved in $a'$ but not in $a''$. Again, since there is a state $s$ where they are tied, $f(s,a')=c' \prec c_1 = f(s,a'')$. 
Thus $a' \succ^i_2 a''$ and therefore $a'$ ordinally dominates $a''$. 
\end{proof}
%
 %

\section{OD and Iterative Voting}
Since ordinal-dominance induces a natural concept of OD-response, we are interested in its implications on iterative voting with multiple strategic voters.
In iterative voting, voters proceed from some initial state $s^{0}$, and in each iteration an arbitrary voter changes her votes, a process that may either converge to an equilibrium or reach a cycle. Our convergence results depend on the structure of the pivot graphs in the epistemic model.

We first show that both cliqued and upward-closed epistemic structures are the result distance-based uncertainty with natural assumptions on the distance function.
\begin{proposition}\label{prop:metric}
\begin{enumerate}
	\item \label{metric_upward} Any neutral distance metric
	$d$ on scoring vectors induces an upward-closed epistemic model.
 \item \label{metric_clique}Any candidate-wise distance metric\footnote{This is a metric on a scoring vector, composed of a singleton metric $D:\mathbb R^2\rightarrow\mathbb [0,1]$, where $d(s,s')=\max_{c\in C}D(s(c),s'(c))$. This includes, for example, the $\ell_{\infty}$ norm.} $d$ on scoring vectors induces a cliqued epistemic model.
\end{enumerate}
\end{proposition}
\begin{proof}[Proof of \ref{metric_upward}]
Assume for contradiction that there is a state $s$ in which there are $c_1,c_{2},c_{3}\in C$ such that $s(c_{1})\geq s(c_{2})\geq s(c_{3})$, $c_2,c_3$ are tied in a state $s'$ within a distance $r$ from $s$, but $c_{1},c_{3}$ are not tied within the same distance. Let $\vec{w}$ be the score vector for $s$, and let $\vec{w'}$ be scoring vector for $s'$. From the triangle inequality, $d(\vec{w}-\vec{w'},0)=d((c'_{1},\ldots,c'_{m}),0)\leq r$. We now examine the $\tilde{\vec{w}}(c'_{2}+w_{1}-w_{2},0,c'_{3},c'_{4},\ldots,c'_{m})$. The first element has to be smaller than $c'_{1}$, and since $\sum_{i=1}^{m}c'_{i}=0$	, we now begin reducing $c'_{4},\ldots,c'_{m}$ until we create a= $\boldsymbol{\tilde{w}'}$, such that its elements sum up to $0$ as well. Since every dimension in the new vector is less than before, $d(\boldsymbol{\tilde{w}'},0)\leq d(\vec{w}-\vec{w'},0)\leq r$, and $\boldsymbol{\tilde{w}'}+\vec{w}$'s has a tie between $c_{1}$ and $c_{3}$.
Proof of \ref{metric_upward}: Assume that there is a state $s=(s_1,\ldots,s_m)$ in which there are $c_1,c_{2},c_{3}\in C$ such that $s_{1}\geq s_{2}\geq s_{3}$; and another state $s'$ within a distance $r$ from $s$ where $c_2,c_3$ are tied. We construct a (non-normalized) vector $s''$ where $c_1,c_3$ are tied, such that $|s''_j-s_j|\leq |s'_j-s_j|$ for all $j$ (hence $s''$ is closer to $s$ than $s'$) or one where $s''$ is such that $s''_{j}=s'_{j}$ for $j>2$ and $|s''_{1}-s_{1}|\leq|s'_{2}-s_{2}|$ and $|s''_{2}-s_{2}|\leq |s'_{1}-s_{1}|$. 

W.l.o.g we have $s'_3\geq s_3$ and $s'_j\leq s_j$ for all $j\neq 2,3$,. Denote by $w=s'_2=s'_3$ be the winning score in $s'$. There are several cases: (I) if $w\geq s_1$, define $s''_1=s_1, s''_2=s_2\leq s''_1, s''_3 = s_1$; (II) if $s_{2}<w<s_1$, define $s''_1 = w\in [s'_1+1,s_1], s''_2 = s_{2}, s''_3=w$ . It is easy to check that $s''$ holds both conditions, thus $d(\frac{s''}{\|s''\|},s)\leq d(s',s)\leq r$ as required. If (III) $w\leq s_{2}$, it quite simple to see that by setting $s''_{1}=w, s''_{2}=w-1,s''_{3}=w$ we are closer to $s$ than $s'$.

Proof of \ref{metric_clique}: any two candidates which are tied with the score leader of $s$ -- $c_{1}$ -- at states at distance $r$ from $s$ are also tied for the leadership in a state within the same distance $r$ from $s$. Since if there is any tie between candidates $c',c''$, either one of them is $c_{1}$, or both of them are tied with $c_{1}$ in the radius $r$ (as the difference in the score of $c_{1}$ in $s$ and the state where it isn't tied for the victory is larger than when it is tied for the win), all candidates which are tied with some other candidate in radius $r$, are tied with $c_{1}$, and hence ``can be tied with $c_1$ in $S_{d,r}(s)$'' is a transitive relation. Since any candidate-wise metric is in particular neutral, $H_{d,r}$ is upward closed by the first part. 

Let $x$ be the lowest-ranked candidate participating in any tie. By upward-closeness, $(y,c_1)\in H_{d,r}$ for all $y$ ranked weakly above $x$. Then by transitivity, any edge $(y,z)$ where $y,z,$ are ranked weakly above $x$ is also in $H_{d,r}$, which means that $H_{d,r}$ is a clique. 
 \end{proof}

Proof of \ref{metric_clique} is similar to Meir~\shortcite{Mei15}, Lemma 2. 
\rmr{As for convergence, a closer look at the convergence proof of local dominance moves from the truthful sate in Plurality~\cite{MLR14}, reveals that it simply exploits the upward-closed structure of the information set, that is induced by the $\ell_1$ metric used in the paper. A somewhat more demanding step is extending convergence proofs from arbitrary states that rely on cliqued structures in Plurality~\cite{Mei15}}

\begin{theorem}
Suppose agents each have a cliqued epistemic model (not necessarily the same one). Iterative voting using Plurality must converge to OD equilibrium, from any initial state. 
%
\end{theorem}
\begin{proof}
For contradiction, let us assume the theorem is wrong and there is a cycle. That is, there is a sequence of scoring vectors (states) ${s}^{1},\ldots {s}^{q}$ such that ${s}^{j+1}$ is the outcome of an agent $i$ making an OD move in ${s}^{j}$, and ${s}^{1}$ is the result of an agent making an OD move in ${s}^{q}$. Let $B$ be the set of candidates whose score changes throughout the cycle, and let $z\in B$ be the candidate with the lowest score in the cycle (if there are multiple such candidates, let $z$ be the lowest ranked in the tie-breaking rule).

Let ${s}^{q'}$ was be a state where $z$ is at their lowest score, and in which an agent $j$ makes a move, changing their vote from some candidate $a$ to $z$. This means $z$ was undominated at this point for $j$, which means all ties with $B$ elements were within the same information set, and moreover, $z\succ^{j}c$ and $c\succ^{j}a$ for any $c\in B$ (Since the pivot-graph is a clique, there is a tie between each 2 candidates in $B$). However, as this is a cycle, there is a step ${s}^{\bar{q}}$, in which agent $j$ changes their vote from $b\in B$ to $a$. This means $b$ is dominated, and $a$ is not, but this means there is some tie between $a$ and another candidate $x$. Since $c\succ^{j} a$ for any $c\in B$, this means $x\notin B$.

If in ${s}^{q'}$ $x$'s score was larger than $z$'s, this means there was a tie between $x$ and $a$ was in the pivot-graph for agent $j$, and by moving to $z$, this indicates $x\succ^{j} a$. If $x$'s score was smaller than $z$'s in ${s}^{q'}$, the score of $b$ in ${s}^{\bar{q}}$ is larger than that of $x$ (since all scores are larger than that of $z$ in ${s}^{q'}$), and since $b\succ_{j}a$, agent $j$ should have preferred to stay with agent $b$. 
\end{proof}

\begin{theorem}\label{itConverge}
Suppose agents each have a concentric, cliqued epistemic model (not necessarily the same one). Iterative voting under \textbf{Veto} must converge to OD equilibrium, from any initial state. 
%
\end{theorem}
\begin{proof}
Assume, for contradiction, that the process does not converge. Let $R$ be the set of candidates whose score changes an infinite number of times, and let $z\in R$ be the candidate which has the lowest score in the cycle (breaking ties using the tie-breaking rule), and let $\vec{s}^{q}$ be the state where it reaches this abysmal score. That is, some voter $j$ moves from vetoing candidate $a$ to vetoing candidate $z$. Candidate $a$'s (and any other $c\in R$) score is above $z$'s, as otherwise its own vetoing before would give it a lower score than $z$. Since this is a cliqued epistemic model, leaving $a$ means it is the favorite candidate of voter $j$ over all candidates with scores above $z$, in particular, for any $c\in R$, $a\succ^{j}c$.

At some point in the future $\vec{s}^{q'}$, due to the cycle, voter $j$ will move from vetoing some candidate $b\in C$ to veto $a$, due to an edge in its relevant pivot-graph, indicating a tie between $a$ and some other candidate $x$. If $x$'s score at $\vec{s}^{q}$ was higher than $z$, then we know $a$ is preferred over it from $z$'s vetoing. If $x$'s score was lower, we know it hasn't changed (as it isn't in $R$), meaning $b$ is still tied with $a$ as well in the pivot-graph of $\vec{s}^{q'}$ as it was in $\vec{s}^{q}$, hence voter $j$ will not move (since $a\succ^{j}b$).
\end{proof}

Other convergence results from \cite{MLR14} could be similarly extended for any upward-closed information structure.
Theorem~\ref{itConverge} with Prop.~\ref{prop:metric} imply the first non-plurality result for local dominance.
\begin{corollary}
Using any candidate-wise metric, local-dominance converges to an equilibrium when using veto.
\end{corollary}

\section{Discussion and Future Directions}

This paper presents a framework to model voting situations in which voters do not have perfect information of the world. Moreover, they do not even have an exact understanding of their uncertainty of the world's state. Hence, their understanding is modeled in a coarser way -- as ``shades of likelihood'' of various voting outcomes, derived from a prospective poll. This framework is robust enough so as to allow us to capture many previously suggested heuristics and strategies of voter behavior under uncertainty. That is, we are able to express these heuristics as rational strategies under particular information structure known to players.

Indeed, the use of the pivot-graph and its topological properties to show convergence (or lack of it), opens the question of whether we can discuss issues of convergence in terms of graph structures (and the metrics or properties that induce them). 
 The fact that ordinal dominance in a large population voting scenario can be computed efficiently, stands in contrast to the negative results in Conitzer et al.~\shortcite{CWX11}, where verifying whether vote $a'$ dominates $a$ is NP-hard under the Borda rule. This is due to our simplifying assumption on the sharp pivot property that allows us to replace (arbitrarily complicated) information sets with a simple pivot graph representation. 

A natural and important use of our model is to reformulate heuristics from various game-theoretic domains -- not limited to social choice -- as ordinally-dominant strategies. This might offer an insight into the built-in assumptions inherent in these heuristics, and allowing, perhaps, novel formulations of new heuristics and methods, tailored to particular uncertainty structures.

Another promising direction is exploring possible connections between ordinal information structures and existing theories of qualitative uncertainty such as \cite{Hal97}.


\section*{Acknowledgments}

This research was partly supported by the Israel Science Foundation (ISF) under Grant No. 773/16 and SUG grant M4082008.020. We acknowledge support from the UK EPSRC under Project EP/P031811/1 (Voting Over Ledger Technologies).

\bibliographystyle{aaai}
\bibliography{general}

\begin{thebibliography}{}

\bibitem[\protect\citeauthoryear{Boutilier}{1994}]{boutilier1994toward}
Boutilier, C.
\newblock 1994.
\newblock Toward a logic for qualitative decision theory.
\newblock {\em KR} 94:75--86.

\bibitem[\protect\citeauthoryear{Brams and Fishburn}{1978}]{brams1978approval}
Brams, S.~J., and Fishburn, P.~C.
\newblock 1978.
\newblock Approval voting.
\newblock {\em American Political Science Review} 72(3):831--847.

\bibitem[\protect\citeauthoryear{Chater, Tenenbaum, and
  Yuille}{2006}]{chater2006probabilistic}
Chater, N.; Tenenbaum, J.~B.; and Yuille, A.
\newblock 2006.
\newblock Probabilistic models of cognition: Conceptual foundations.
\newblock {\em Trends in Cognitive Sciences} 10(7):287--291.

\bibitem[\protect\citeauthoryear{Conitzer, Walsh, and Xia}{2011}]{CWX11}
Conitzer, V.; Walsh, T.; and Xia, L.
\newblock 2011.
\newblock Dominating manipulations in voting with partial information.
\newblock In {\em Proceedings of the 25th National Conference on Artificial
  Intelligence (AAAI)},  638--643.

\bibitem[\protect\citeauthoryear{Desmedt and Elkind}{2010}]{DE10}
Desmedt, Y., and Elkind, E.
\newblock 2010.
\newblock Equilibria of plurality voting with abstentions.
\newblock In {\em Proceedings of the 11th ACM conference on Electronic Commerce
  (EC)},  347--356.

\bibitem[\protect\citeauthoryear{Dow and Werlang}{1994}]{dow1994nash}
Dow, J., and Werlang, S. R. d.~C.
\newblock 1994.
\newblock Nash equilibrium under knightian uncertainty: breaking down backward
  induction.
\newblock {\em Journal of Economic Theory} 64(2):305--324.

\bibitem[\protect\citeauthoryear{Dubois, Fargier, and Prade}{1996}]{DFP96}
Dubois, D.; Fargier, H.; and Prade, H.
\newblock 1996.
\newblock Refinements of the maximin approach to decision-making in a fuzzy
  environment.
\newblock {\em Fuzzy Sets and Systems} 81(1):103--122.

\bibitem[\protect\citeauthoryear{Dutta and Laslier}{2010}]{DL10}
Dutta, B., and Laslier, J.-F.
\newblock 2010.
\newblock Costless honesty in voting.
\newblock in 10th International Meeting of the Society for Social Choice and
  Welfare, Moscow.

\bibitem[\protect\citeauthoryear{Elkind \bgroup et al\mbox.\egroup
  }{2015}]{elkind2015equilibria}
Elkind, E.; Markakis, E.; Obraztsova, S.; and Skowron, P.
\newblock 2015.
\newblock Equilibria of plurality voting: Lazy and truth-biased voters.
\newblock In {\em International Symposium on Algorithmic Game Theory},
  110--122.
\newblock Springer.

\bibitem[\protect\citeauthoryear{Grandi \bgroup et al\mbox.\egroup
  }{2013}]{GLRVW13}
Grandi, U.; Loreggia, A.; Rossi, F.; Venable, K.~B.; and Walsh, T.
\newblock 2013.
\newblock Restricted manipulation in iterative voting: {C}ondorcet efficiency
  and {B}orda score.
\newblock In {\em Proceedings of 3rd International Conference of Algorithmic
  Decision Theory (ADT)},  181--192.

\bibitem[\protect\citeauthoryear{Halpern}{1997}]{Hal97}
Halpern, J.~Y.
\newblock 1997.
\newblock Defining relative likelihood in partially-ordered preferential
  structures.
\newblock {\em Journal of Arti􏰀cial Intelligence Research} 7:1--24.

\bibitem[\protect\citeauthoryear{Laslier}{2009}]{Las09}
Laslier, J.-F.
\newblock 2009.
\newblock The leader rule: A model of strategic approval voting in a large
  electorate.
\newblock {\em Journal of Theoretical Politics} 21(1):113--136.

\bibitem[\protect\citeauthoryear{Laslier}{2010}]{laslier2010laboratory}
Laslier, J.-F.
\newblock 2010.
\newblock Laboratory experiments on approval voting.
\newblock In {\em Handbook on approval voting}. Springer.
\newblock  339--356.

\bibitem[\protect\citeauthoryear{Matt, Toni, and
  Vaccari}{2009}]{matt2009dominant}
Matt, P.-A.; Toni, F.; and Vaccari, J.~R.
\newblock 2009.
\newblock Dominant decisions by argumentation agents.
\newblock In {\em International Workshop on Argumentation in Multi-Agent
  Systems},  42--59.
\newblock Springer.

\bibitem[\protect\citeauthoryear{Meir, Lev, and Rosenschein}{2014}]{MLR14}
Meir, R.; Lev, O.; and Rosenschein, J.~S.
\newblock 2014.
\newblock A local-dominance theory of voting equilibria.
\newblock In {\em Proceedings of the 15th ACM conference on Economics and
  Computation (EC)},  313--330.

\bibitem[\protect\citeauthoryear{Meir}{2015}]{Mei15}
Meir, R.
\newblock 2015.
\newblock Plurality voting under uncertainty.
\newblock In {\em Proceedings of the 29th Conference on Artificial Intelligence
  (AAAI)},  2103--2109.

\bibitem[\protect\citeauthoryear{Meir}{2017}]{ME17}
Meir, R.
\newblock 2017.
\newblock Iterative voting.
\newblock In Endriss, U., ed., {\em Trends in Computational Social Choice}. AI
  Access.
\newblock chapter~4,  69--86.

\bibitem[\protect\citeauthoryear{Myerson and Weber}{1993}]{MW93}
Myerson, R.~B., and Weber, R.~J.
\newblock 1993.
\newblock A theory of voting equilibria.
\newblock {\em The American Political Science Review} 87(1):102--114.

\bibitem[\protect\citeauthoryear{Reijngoud and Endriss}{2012}]{RE12}
Reijngoud, A., and Endriss, U.
\newblock 2012.
\newblock Voter response to iterated poll information.
\newblock In {\em Proceedings of the 11th International Conference on
  Autonomous Agents and Multiagent Systems (AAMAS)}, volume~2,  635--644.

\bibitem[\protect\citeauthoryear{Shafer}{1976}]{shafer1976mathematical}
Shafer, G.
\newblock 1976.
\newblock {\em A mathematical theory of evidence}, volume~42.
\newblock Princeton university press.

\bibitem[\protect\citeauthoryear{Tversky and Kahneman}{1974}]{TK74}
Tversky, A., and Kahneman, D.
\newblock 1974.
\newblock Judgment under uncertainty: Heuristics and biases.
\newblock {\em Science} 185(4157):1124--1131.

\bibitem[\protect\citeauthoryear{Williamson}{1992}]{williamson1992inexact}
Williamson, T.
\newblock 1992.
\newblock Inexact knowledge.
\newblock {\em Mind} 101(402):217--242.

\end{thebibliography}


%
%
%
%
%
\end{document}